# On a Positron Channeling in Ionic Crystals

E.A. Ayryan [1] , A.S. Gevorkyan [2,3] , M. Hnatic[4,5,6] ,

K.B. Oganesyan [1,7*] , P. Kopcansky [6], Yu.V. Rostovtsev [8] , M. Timko [6]

[*] bsk@yerphi.am

[1] LIT, Joint Institute for Nuclear Research, Dubna, Russia
[2] Institute for Informatics and Automation Problems,
[3] Institute for Chemical Physics, NAS of Armenia,Yerevan, Armenia,
[4] BLTP, Joint Institute for Nuclear Research, Dubna, Russia

[5] Faculty of Sciences, P. J. Safarik University, Kosice, Slovakia

[6] Institute of Experimental Physics SAS, Kosice, Slovakia

[7] A.I. Alikhanyan National Science Lab, Yerevan Physics Institute, Yerevan, Armenia,

[8] University of North Texas, Denton, TX, USA,

An analytical expression is received for the effective interaction potential of a fast charged particle with the ionic crystal CsCl near the direction of axis <100> as a function of the temperature of the medium. A possibility of positron channeling in this potential is shown. By numerical analysis it is shown that the effective potential of axial channeling of positrons along the axis <100> of negatively charged $Cl^{-}$ ions practically does not depend on temperature of the media.

## 1. Introduction

The phenomenon of anomalous passage of ions along definite crystallographic axes and planes was discovered experimentally in 1960 [1]. In 1963 it was confirmed by numerical simulation [2] and became known as the channeling effect. In 1965 Lindhard theoretically explained this phenomenon within the limits of classical mechanics [3]. The quantum theory of channeling for electrons and positrons has been elaborated by many authors [4–6]. It is important to note that an electron in a crystal can commit both planar and axial channeling. At the same time only one type of pure channeling for the positrons is known, the regime where a particle is localized between two adjacent planes. The possibility of axial channeling of positive particles has not been investigated, seriously up to now, because the crystallographic axes, irrespective of grade of crystal, have been charged positively.

However, investigation of possibilities of axial channeling of positrons and, hence the formation of metastable relativistic positron systems is a problem of utmost importance for radiation physics. At first, for the solution of this problem, the method of creation of positronium (Ps) atoms was proposed with the help of mixing of beams of relativistic electrons and positrons in vacuum with the subsequent stimulation of decay of a Ps atom on two photons ( $e^+e^- \to 2\gamma$ [7,8]. Another development route was Ps atom creation by a photon in a field of atoms [9] or Coulomb fields [10,11]. However, this method is virtually ineffective for applications, because the probability of Ps atom formation by this method is extremely small. Up to now different methods have been proposed for generation of coherent $\gamma$ rays based on the effect of channeled positron annihilation with the electrons of media [12-16]. We believe that this direction of investigation is more preferable. In [17-20] the channeling of electrons in an intense light wave was discussed.

In an earlier study we focused our attention on the capabilities of ionic crystals of the CsCl type, with the aim of finding new possibilities for the channeling of light charged particles. In particular, we studied in detail the effective interaction potential of a charged particle with a crystal under conditions of planar channeling along the principal <100> planes of cesium $Cs^+$ and chlorine $Cl^-$ ions. In the present paper we construct the effective interaction potential of a charged relativistic particle with a crystal near the direction of axis <100>.

In the present paper the effective interaction potential of the positron with crystal near the <100> axis of the $Cl^-$ ions is investigated in detail by numerical simulation. It is shown that the dependence of the effective potential for the positron from the temperature of the medium is very weak.

## 2. Axial channeling of positrons along axes <100> of chlorine ions $Cl^-$

The electrostatic potential which creates the infinite crystal of type CsCl along the axes <100> of negative chlorine ions $Cl^-$ in the finite temperature $T \neq 0$ may be written in the following form [14]:

$$\varphi(\mathbf{r},T) = \frac{4\pi}{d^3} \int d\mathbf{R} \sum_{k \neq 0} \frac{\exp[i\mathbf{k}(\mathbf{r}-\mathbf{R})}{\mathbf{k}^2} \left\{ \Lambda_-^{(k)} W^-(\mathbf{R}) + (-1)^{l+m+n} \Lambda_+^{(k)} W^+(\mathbf{R}) \right\},$$
$$\Lambda_\pm^{(k)} = \exp\left( -\frac{k^2 u_{0\pm}^2}{2} \right), \tag{1}$$

where d is a lattice constant, $\mathbf{k} = (2\pi / d)(\hat{x}l + \hat{y}n + \hat{z}m)\mathrm{k}$, $k = |\mathbf{k}|$ is the reciprocal lattice vector, (l, n, m) $\in (-\infty, +\infty)$ are integers, $u_{0-}(T)$ and $u_{0+}(T)$ characterize the amplitudes of the thermal vibrations of the positive $Cs^{+}$ and the negative $Cl^{-}$ ions, $W^{-}(\mathbf{R})$ and $W^{+}(\mathbf{R})$ correspondingly denote the charge density in ions at temperature T=0. The electrostatic field (1) is greatly simplified outside the ionic lattice, and assumes the form:

$$\varphi_{ost}(\mathbf{r};T) = \frac{4\pi e^{-}}{d^{3}} \sum_{k \neq 0} \frac{\exp(i\mathbf{kr})}{\mathbf{k}^{2}} \left\{ \Lambda_{-}^{(k)} - (-1)^{l+m+n} \Lambda_{+}^{(k)} \right\}. \tag{2}$$

Now we can transfer to the problem of investigating the effective potential. Let a fast, positively charged particle scatter under a small corner $\vartheta < \vartheta_{L} \sim (D_{0} / E)^{1/2}$ on the <100> axis of $Cl^{-}$ ions, where $\vartheta_{L}$ is the Lindhard angle, $E$ is the total energy of the particle, and $D_{0}$ is the depth of the well. The potential (1) can then be averaged along the direction of fast motion, i.e. along the <100> axis of $Cl^{-}$ ions, which is equivalent to the integration of the potential along the coordinate z within the limits of one period d (Figure 1). If the particle crosses a $Cl^{-}$ ion at a distance $\rho = \sqrt{x^{2} + y^{2}}$ , the period d is divided into three parts (Figure 1(c)). In the first and third segments of d the particle is moving outside the ion, while in the second segment it moves inside the ion.

Note that the length of the path crossed by the particle inside the ion is:

$$\begin{gathered} R_{-}(x,y) = 2\,\mathrm{Re}\left( R_{0-}^{2} - \left[ \eta^{2}(x) + \eta^{2}(y) \right] d^{2} \right)^{1/2}, \\ \eta(u) = \frac{1}{2} + (-1)^{p} \left( \left\{ \frac{u}{d} \right\} - \frac{1}{2} \right), \end{gathered} \tag{3}$$

where the brackets [.] and {.} denote the integer and the fractional parts of a function, respectively. The crossed path in the ion $Cs^{+}$ is given by the formula:

$$R_{+}(x,y) = 2\,\mathrm{Re}\left\{ R_{0+}^{2} - d^{2} \left[ \eta^{2}\left( x - \frac{d}{2} \right) + \eta^{2}\left( x - \frac{d}{2} \right) \right] \right\}^{1/2}. \tag{4}$$

Recall that in Equations (3) and (4) the symbols $R_{0+}$ and $R_{0-}$ denote the radii of the corresponding ions. The average potential may be written in the form:

$$\varphi_{eff}(x,y;T) = \left\{ \int_{-R_{-}/2}^{R_{-}/2} - \int_{-R_{+}/2}^{R_{+}/2} \varphi(x,y,z;T)\,dz \right\} + \left\{ \int_{-R_{-}/2}^{R_{-}/2} + \int_{-R_{+}/2}^{R_{+}/2} \varphi(x,y,z;T)\,dz \right\}. \tag{5}$$

Substituting expressions (1) and (2) into (5) and after elementary integration we find:

$$
\begin{aligned}
&\varphi_{eff}(x,y;T) = \frac{8|e^-|}{\pi^2 d} \sum_{\substack{l,m,n=0\\ l+m+n>0}} a_l a_m a_n \frac{\exp(-\lambda^2\mu^2)}{m\mu^2} \cos(k_l x)\cos(k_n y) \\
&\times\left\{(-1)^{l+n}\sin\left[\frac{1}{2}k_m R_+(x,y)\right]W^+(l,m,n) + \sin\left[\frac{1}{2}k_m R_-(x,y)\right]W^-(l,m,n)\right\} \\
&+\frac{4|e^-|}{\pi d}\sum_{\substack{l,n=0\\ l+n>0}} a_l a_n \frac{\exp(-\lambda^2\upsilon^2)}{\upsilon^2}\cos(k_l x)\cos(k_n y)\left\{(-1)^{l+n}-1\right\}, \\
&k_q = \frac{2\pi}{d}q,
\end{aligned}
\tag{6}
$$

where the following notations are introduced: $\mu^2 = l^2 + n^2 + m^2$, $\upsilon = \mu(m=0)$, $\lambda = u_{0\pm}/d$, $a_0 = 1/2$ and $a_i = 1(i \neq 0)$, in addition:

$$
W^{\pm}(l,m,n) = \int W^{\pm}(\mathbf{R})\exp(-i\mathbf{kR})d\mathbf{R} \tag{7}
$$

Note that the parameter $\lambda$ is obtained on the assumption that the thermal vibration amplitudes are equal to $u_{0+} = u_{0-}$, which is a good approximation within the model of acoustic vibrations. The form factors $W^+(l,m,n)$ and $W^-(l,m,n)$ may be simplified for the numerical analysis. For the charge density in the ions, we can write this in the form:

$$
W^{\pm}(\mathbf{R}) = V^{\pm}(\mathbf{R}) + Z^{\pm}\delta(\mathbf{R}), \tag{8}
$$

where $V^{\pm}(\mathbf{R})$ is the distribution of electrons inside ions of the crystal in the Jensen–Mayer–Gosler–Rode approximation (JMGR) and $Z^{\pm}$ is the number of protons in the nucleus [21]. Substituting expression (8) into (7) and assuming that the distribution of the electron charge inside the ion has spherical symmetry, we obtain the following expression for the structure factor:

$$
\begin{aligned}
&W^{\pm}(l,m,n) = Z^{\pm} \pm 1 + X^{\pm}(l,m,n) \\
&X^{\pm}(l,m,n) = \frac{4\pi}{k}\int V^{\pm}(R) R \sin(kR) dR
\end{aligned}
\tag{9}
$$

Further, using for the function V(R) the Lenz– Jensen model with the parameters of the crystal CsCl, expression (9) is calculated for four different values of $\lambda$ ($\lambda$=0.001, 0.01, 0.05 and 0.1). As one can see from Figures 2(a)–(d), for positively charged fast particles around the <100> axis of $Cl^-$ ions, a rather broad potential exists (width $\Delta d \sim 0.25d$) with the depth of order ~10 eV, which remains constant for a large range of thermal vibrations 0.001–0.1, i.e. for a broad range of temperatures. In other words, the relativistic positively charged particle at the scattering under the small angle on the <100> axis of the $Cl^-$ ions can be captured in the regime of axial channeling during which the scattering of a positron on the phonon's subsystem does not occur. Based on the symmetry of the resulting effective potential (Figures 2(a)–(d)), the effective potential conveniently may be fitted by the function:

$$U_0(\rho) = D_0(\mathrm{e}^{-2\alpha\bar{\rho}} - 2\,\mathrm{e}^{-\alpha\bar{\rho}}), \quad \bar{\rho} = \frac{\rho - \rho_0}{\rho_0}, \tag{10}$$

where potential (10) has the following parameters: $D_0$ = 9.8 eV, $\alpha$ = 0.838 and $\rho_0$ = 0.46d. It is important to note that the potential (10), as we have shown, well enough approximates the exact effective field (6). It is enough to say, that in the regions where the potential U< -4 eV, the approximation error is below 1%. Finally we want to note that for the numerical calculation of expressions (9) the parameters of ions $Cl^-$ and $Cs^+$ in a free state are used. For this reason the characteristics of the computed effective potential for the axial channeling of positrons may be partially different from reality in the impairment direction.

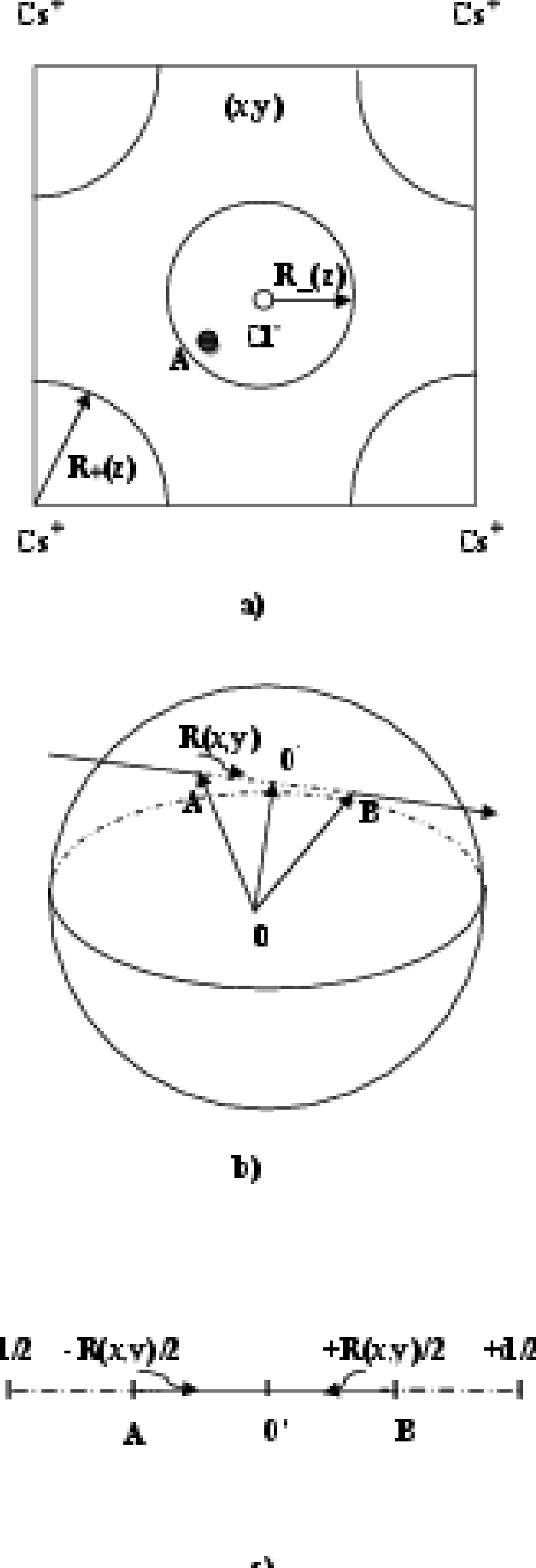

Figure 1. (a) Two-dimensional cross-section {x, y} of the three-dimensional cell of CsCl crystal at depth z along the <100> axis. The radii of the cross-sections of the ion spheres are defined by the expressions $R_+(z) = \mathrm{Re}[R_{0+}^2 - (d/2 - z)^2]^{1/2}$ and $R_-(z) = \mathrm{Re}[R_{0-}^2 - z^2]^{1/2}$;
(b) intersection of the particle path with the sphere of a lattice ion at points A and B; (c) the hatched part indicates the path segment of length R(x, y), traversed by the particle inside the ion.

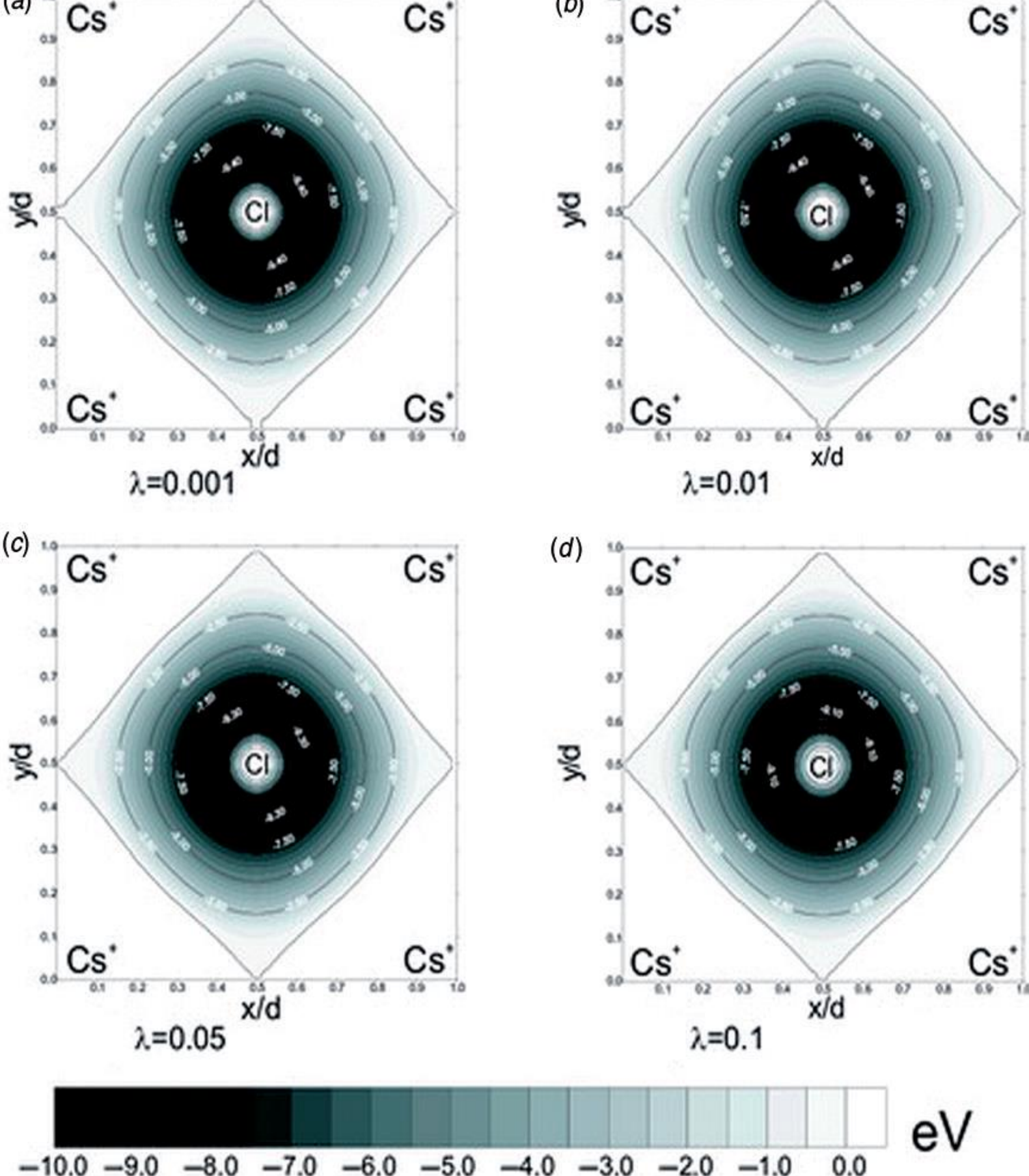

Figure 2. Profile of the effective potential for the axial channeling of a positron along the <100> axis of $Cl^-$ ions at various temperatures.

## 3. Conclusion

The possibility of positron channeling in ionic crystals is investigated. As was shown by numerical simulation the channeling potential for a positron has annular symmetry in the considered case and is situated in regions far from the crystal axes, and is extremely weakly dependent on the temperature of the media.

**Acknowledgment**

For EAA and KBO, this work was supported by the NSP of Slovak Republic.